\documentclass[aps,twocolumn,superscriptaddress,showpacs]{revtex4}
\usepackage{amsfonts}
\usepackage{amsmath}
\usepackage{amssymb}
\usepackage{graphicx}
\usepackage{color}

\begin{document}
\title{Electromagnetically Induced Transparency of Interacting Rydberg Atoms
	with Two-Body dephasing}
\author{Dong Yan}
\email{ydbest@126.com}
\affiliation{School of Science and Key Laboratory of Materials Design and Quantum
	Simulation, Changchun University, Changchun 130022, P. R. China}
\affiliation{School of Physics and Astronomy, University of Nottingham, Nottingham, NG7
	2RD, United Kingdom}
\author{Binbin Wang}
\affiliation{School of Science and Key Laboratory of Materials Design and Quantum
	Simulation, Changchun University, Changchun 130022, P. R. China}
\author{Zhengyang Bai}
\affiliation{School of Physics and Astronomy, University of Nottingham, Nottingham, NG7
	2RD, United Kingdom}
\affiliation{State Key Laboratory of Precision Spectroscopy, East China Normal
	University, Shanghai 200062, P. R. China}
\author{Weibin Li}
\affiliation{School of Physics and Astronomy, University of Nottingham, Nottingham, NG7
	2RD, United Kingdom}
\date{\today }

\begin{abstract}
We study electromagnetically induced transparency of a ladder type
configuration in ultracold atomic gases, where the upper level is an
electronically highly excited Rydberg state. The strong two-body interaction in the Rydberg state leads to the excitation blockade, where all but one atoms are shifted out of resonance such that the transmission of the probe light is affected. We show that  molecular coupling in the Rydberg state causes an effective, two-body dephasing. The presence of the two-body dephasing leads to a similar blockade effect. Hence the overall blockade effect is enhanced by the two-body dephasing.~Through numerical and approximately analytical calculations, we find that the transmission is reduced drastically by the presence of two-body dephasing in the transparent
window, which is accompanied by strong photon-photon anti-bunching. Around
the Autler-Townes splitting, the photon bunching is amplified by the two-body dephasing.
\end{abstract}

\maketitle

\section{Introduction}

Electromagnetically induced transparency (EIT)~\cite
{Harris,Boller,Fleischhauer} plays a pivotal role in
quantum and nonlinear optics~\cite
{Liu,Lukin,Chaneliere,Appel,Lvovsky} and has been investigated intensively in the
past two decades~
\cite{Paternostro,Payne,Ottaviani,Petrosyan1}. Recently
there has been a growing interest in the study of EIT using
electronically highly excited (Rydberg) states with principal quantum number $n\gg 1$.
Rydberg atoms have long life times ($\sim n^3$) and strong two-body interactions (e.
g. van der Waals interaction strength $\sim n^{11}$). The distance dependent interaction can suppress multiple Rydberg excitation of nearby atoms, giving rise to the so-called Rydberg excitation blockade. By mapping the Rydberg atom interaction to light fields
through EIT~\cite{Gorshkov1}, strong and long-range interactions between
individual photons can be achieved. This permits to study nonlinear
quantum optics at the few-photon level~\cite{Firstenberg1,Liang} and find
quantum information applications~\cite{Saffman2} to create single photon sources~\cite%
{Saffman1,Walker,Muller}, filters~\cite{Peyronel,Gorshkov1},
substractors~\cite{Honer,Gorshkov2}, transistors~%
\cite{Iiarks,Gorniaczyk1}, switches~\cite{Chen,Baur}, and
 gates~\cite{Friedler,Paredes}.

On the other hand, dephasing and decay of Rydberg atoms are unavoidable due to, e.g., atomic motions and finite lasers linewidth~\cite{Scully}. In the study of long time dynamics, it has been shown that dissipation of individual atoms competes the two-body
Rydberg interactions as well as laser-atom coupling. The interplay leads to interesting
driven-dissipative many-body dynamics, such as glassy behaviors induced by
single atom dephasing~\cite{Lesanovsky},
bistability and metastability~\cite{Letscher,Macieszczak}, Mott-superfluid phase
transition~\cite{Ray}, emergence of antiferromagnetic phases~\cite{Hoening},
dissipation controlled excitation statistics~\cite{Schonleber}, and
dissipation induced blockade and anti-blockade~\cite{Young}. Nonetheless, collective dissipative processes emerge in dense atomic gases, typically through two-body dipolar couplings~\cite{Ficeka02}. Well-known examples are the sub- and super-radiance in atomic ensembles. In cold atom gases, energies of pairs of atoms in different Rydberg states can be close, leading to strong dipolar interaction through the F\"orster resonance~\cite{zeros05,nipper12,ravets14,anto16,Asaf16,Gorniaczyk2}. Due to the coupling between different atomic pair (molecular) states, strong dephasing was observed in the study of Rydberg-EIT~\cite{Tresp}.

In this work, we study Rydberg-EIT with a model where both van der Waals interactions and two-body dephasing  are present. The latter could be induced by molecular transitions due to the presence of multiple Rydberg states~\cite{Macieszczak,Gorniaczyk2,Li,Cano,Liu2,Huang,Viscor}. 
Starting from the dipolar coupling between Rydberg pair states, we derive a master equation in which van der Waals (vdW) interactions and effective two-body dephasing (TBD) are both present in the target Rydberg states. Using a superatom (SA) method~\cite{Petrosyan2}, we study stationary properties of the Rydberg-EIT~\cite{Weatherill,Pritchard1,Pritchard2,Reslen,Ates,Yan1,Yan2,Jen,Garttner1,Stanojevic} due to the interplay between the coherent and incoherent two-body processes. We find that the blockade radius is enlarged by the two-body dephasing. As a result, the transmission
and photon-photon correlation of the probe field is modified by the dephasing in and out-of the EIT window.
\begin{figure}[htbp]
	\centering
	\fbox{\includegraphics[width=0.95\linewidth]{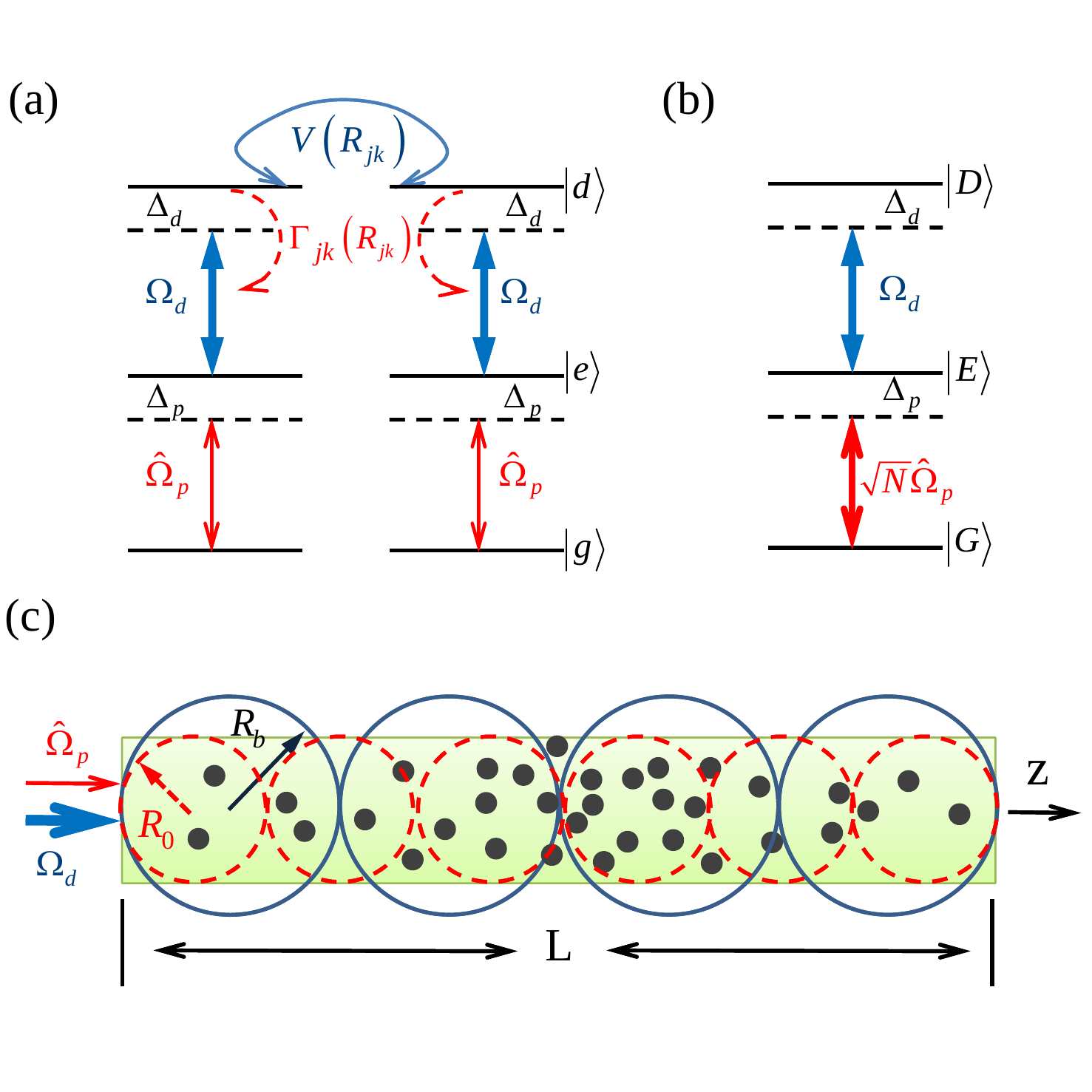}}
	\caption{(Color online) (\textbf{a}) Atomic levels. A weak probe field (Rabi frequency $\Omega_{p}$ and
		detuning $\Delta_p$) and a classical coupling field (Rabi frequency $\Omega _{d}$ and detuning $%
		\Delta_d$) couple the ground stat $|g\rangle $, intermediately state $|e\rangle $ and Rydberg state $|d\rangle $, respectively. $V_{jk}$ and $%
		\Gamma_{jk}$ are long-range van der Waals interaction and
		two-body dephasing, which depend on atomic separation $R_{jk}$. (\textbf{b}) The equivalent superatom (SA) which is composed of three collective state $|G\rangle $, $|E\rangle $ and $%
		|D\rangle $. The coupling between the collective state $|G\rangle$ and $|E\rangle$ is enhanced by a factor of $%
		\protect\sqrt{N}$. (\textbf{c}) The SAs with (blue solid line) and without (red dashed line)
		TBD in a quasi one-dimensional atomic ensemble (length $L$). The number of SAs decreases as the size of the SAs increases.}
	\label{fig1}
\end{figure}

The structure of the paper is as follows. In Sec. II, the Hamiltonian and many-body master equation that is capable to capture the two-body processes is introduced.
 In Sec. III, the modification of the blockade radius by the two-body dephasing is discussed. This is done by numerically solving the master equation for two atoms and analytically through an effective Hamiltonian. In Sec. IV, we solve the light propagation and atomic dynamics through the Heisenberg-Langevin approach. We identify
parameters where the transmission of the probe light is affected most by the TBD. We show that the photon-photon correlations are drastically modified by the TBD in the transparent window and around Autler-Townes splitting. We conclude in Sec. VI.

\section{Many-atom master equation model}
We consider a cold gas of $N$ Rb atoms, which are described by a
three-level ladder type configuration with a long-lived ground state $%
|g\rangle $, a low-lying excited state $|e\rangle $ with decay rate $%
\gamma_e$, and a highly excited Rydberg state $|d\rangle$. The level
scheme is shown in Fig.\ref{fig1}(a). Specifically these states are given by $|g\rangle=|5S\rangle $, $|e\rangle
=|5P\rangle $ and $|d\rangle =|nD\rangle$. The upper transition$\,|e\rangle
\rightarrow \,|d\rangle $ is driven by a classical control field with Rabi
frequency $\Omega _{d}$ and detuning $\Delta _{d}$. The lower transition $%
\,|g\rangle \rightarrow \,|e\rangle $ is coupled by a weak laser field,
whose electric field operator and detuning is given by $\hat{\mathcal{E}}%
_{p} $ and $\Delta _{p}$, respectively. In Rydberg states, two atoms located
at $\mathbf{r}_{j}$ and $\mathbf{r}_{k}$ will experience long-range
 van der Waals (vdW) type interaction,
given by $V_{jk} =\hbar C_{6}/R_{jk}^{6}$, where $C_{6}$ is
the dispersion coefficient and\ $R_{jk}=\left\vert \mathbf{r}_{j}-\mathbf{r}%
_{k}\right\vert $ is the distance between the two atoms. The Hamiltonian of
the system reads
\begin{equation}
\hat{H}=\hat{H}_{0}+\hat{V}_{d}\left( R\right) ,  \label{Ham}
\end{equation}%
where $\hat{H}_{0}=\sum_{j=1}^N[\Delta _{p}\hat{\sigma}_{ee}^{j}+\left( \Delta
_{p}+\Delta _{d}\right) \hat{\sigma}_{dd}^{j}]+[\hat{\Omega}_{p}\hat{\sigma}%
_{eg}^{j}+\Omega _{d}\hat{\sigma}_{ed}^{j}+\text{H.c}]$ describes the
atom-light interaction. We have defined the Rabi frequency operator $\hat{%
\Omega}_p = g\hat{\mathcal{E}}_{p}$ with $g$ the single atom coupling
constant~\cite{Scully}. $\hat{V}_{d}\left( R\right) =\sum_{j>k}V_{jk} \hat{\sigma}_{dd}^{j}\hat{\sigma}_{dd}^{k}$ is the vdW
interaction between Rydberg atoms. Here $\hat{\sigma}_{mn}^{j}=|m\rangle_j%
\langle n|$ is the transition operator of the $j$-th atom.

A pair of atoms in the Rydberg $|d\rangle$ state can couple to other pairing states of similar energies, due to the small quantum defects in Rydberg $|d\rangle$ state as well as the presence of F\"oster resonances. The form of this coupling is assumed to be the dipole-dipole interaction. Typically we will have to deal with a large number of such pair states, which make the analysis extremely challenging. In this work, we will treat these background molecular pairs perturbatively. We will assume atoms in the background molecular states decay rapidly to the $|d\rangle$ state. This allows us to adiabatically eliminate the molecular states, which leads to an effective, two-body dephasing in the $|d\rangle$ state (see Appendix for derivation). Further taking into account of other decay processes, dynamics of the many-atom system is governed by the following master equation
\begin{eqnarray}
\dot{\varrho} &=&-i[\hat{H},\varrho]+2\gamma _{e}\sum_{j}\left(\hat{\sigma}%
_{ge}^{j}\varrho \hat{\sigma}_{eg}^{j}-\frac{1}{2}\{\varrho ,\hat{\sigma}_{eg}^{j}%
\hat{\sigma}_{ge}^{j}\}\right)  \notag \\
&&+2\gamma _{d}\sum_{j}\left(\hat{\sigma}_{dd}^{j}\varrho \hat{\sigma}_{dd}^{j}-%
\frac{1}{2}\{\varrho ,\hat{\sigma}_{dd}^{j}\}\right)  \notag \\
&&+\sum_{j>k}{\Gamma}_{jk}\left( \hat{\sigma}_{dd}^{j}\hat{\sigma}%
_{dd}^{k}\varrho \hat{\sigma}_{dd}^{k}\hat{\sigma}_{dd}^{j}-\frac{1}{2}\{\varrho,%
\hat{\sigma}_{dd}^{k}\hat{\sigma}_{dd}^{j}\}\right),
\label{MasterEqu}
\end{eqnarray}%
where $\gamma_d$ is single atom dephasing rate in state $|d\rangle$. The two-body dephasing ${\Gamma}_{jk}=\hbar {\Gamma}_{6}/R_{jk}^{6}$
with ${\Gamma}_{6}$ being a coefficient characterizing the strength of the TBD. 

\section{Two-body dephasing enhanced blockade effect}
\label{sec:blockade}
In this section, we reveal main effects caused by the two-body dephasing with a simple example of two atoms. We first calculate stationary states of two atoms by solving the master equation (\ref{MasterEqu}) numerically. Using the stationary state solution, we evaluate the two-body correlation
\begin{equation}
C(R_{12})=\frac{\langle \hat{\sigma}_{dd}^1\hat{\sigma}_{dd}^2\rangle}{\langle\hat{\sigma}_{dd}^1\rangle\langle\hat{\sigma}_{dd}^2\rangle}.
\end{equation}

\begin{figure}[htbp]
	\centering
	\fbox{\includegraphics[width=0.95\linewidth]{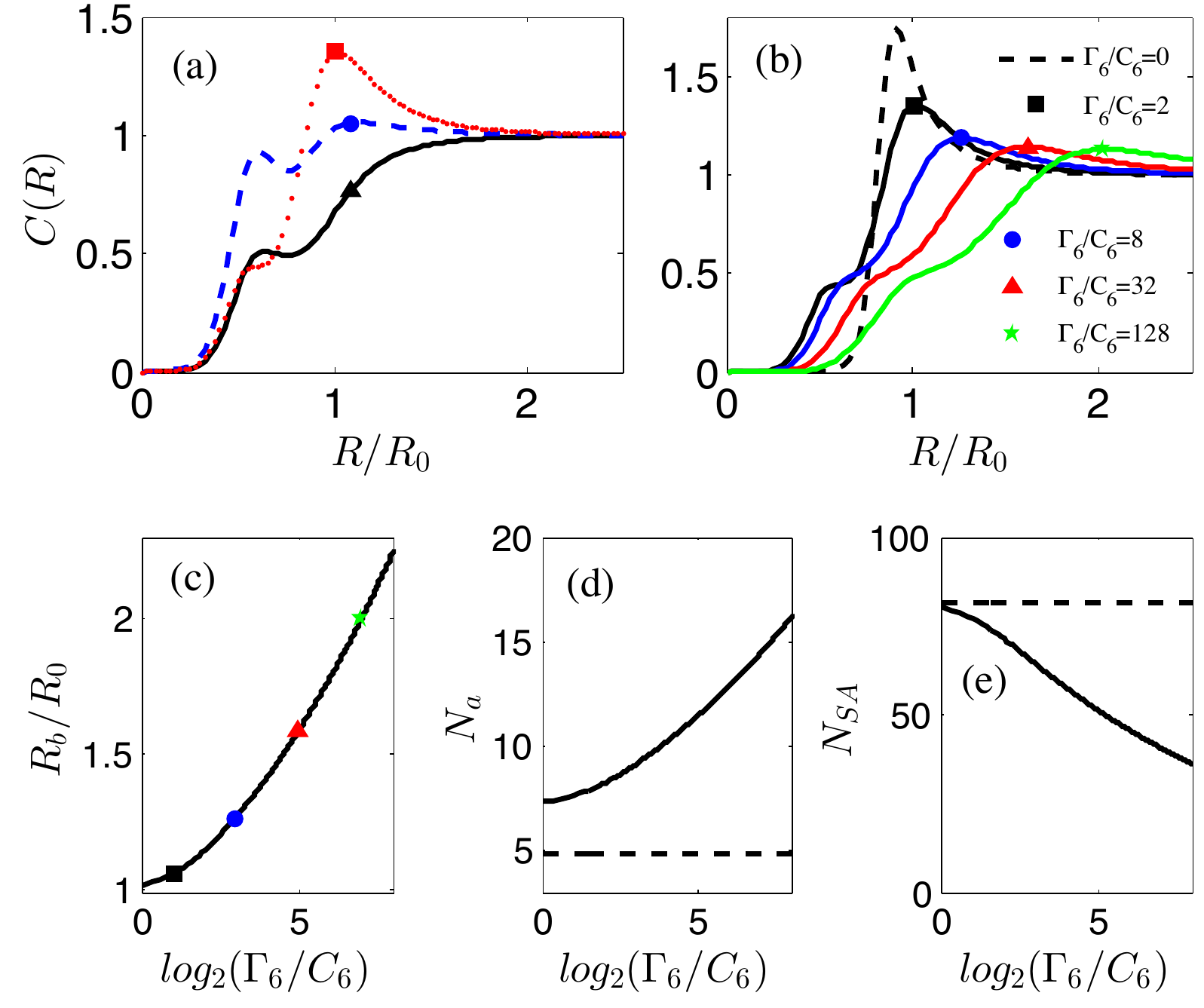}}
	\caption{(Color online) (\textbf{a}) Correlation function $C(R)$ for $\Delta_d=-\Delta_p=-0.3$ MHz (solid), $
		\Delta_d=-\Delta_p=-2.0$ MHz (dashed) and $\Delta_d=-\Delta_p=-4.0$ MHz (dotted). A maximum is found when the single photon detuning ($|\Delta_p|=|\Delta_p|$) is large. Other parameters are ${\Gamma}_{6}=2C_{6}$ and $\Omega _{p}/2\protect\pi =0.5$ MHz. (\textbf{b}) Correlation function $C(R)$ with large single photon detuning $\Delta_d=-\Delta_p=-4.0$ MHz.
		Increasing the TBD rate $\Gamma _{6}$, the maximal values reduce
		gradually. (\textbf{c}) $R_{b}$ v.s. $\Gamma_6$. The location corresponding to the maximal value of the correlation function is marked (see Fig.~\ref{fig:correlation}b). (\textbf{d}) Number $N_{a}$ of atoms per
		superatom and (\textbf{e}) number $N_{SA}$ of superatoms in the
		one-dimensional atomic ensemble. As the blockade radius increases with $\Gamma_6$, the volume of a superatom become larger. Fixing the length of the medium, total number of superatoms is reduced. Other parameters for panels (b-e) are $\Omega _{d}/2 \protect%
		\pi =2.0$ MHz, $\protect\gamma _{e}/2%
		\protect\pi =3.0$ MHz, $\protect\gamma _{d}/2\protect\pi =10.0$ kHz, $C_{6}/2%
		\protect\pi=140 \,\text{GHz}\, \mu \text{m}^6$, and $L=1.0$ mm.}
	\label{fig:correlation}
\end{figure}

Under the two-photon resonance condition $\Delta_{p}+\Delta_{d}=0$, the two-body correlation is shown in Fig.~\ref{fig:correlation}a. At short distances, simultaneous excitation of the two atoms is prohibited, where $C(R)$ is small. Increasing the atomic distance $R$, the correlation increases, and saturates at $C\sim 1$ at large distances, i.e. independent excitation of Rydberg atoms. For intermediate distances, the correlation function displays features strongly depending on the single photon detuning.

For sufficiently large single photon detuning $|\Delta_{p}|=|\Delta_{d}|\gg |\Omega_p|$, the correlation shows a maximum, as can be seen in Fig.~\ref{fig:correlation}a. In this case, the correlation increases at short distances and decrease at large distances with increasing $\Gamma_6$. The maximal value of the correlation decreases with increasing TBD rate $\Gamma_6$ (see Fig.~\ref{fig:correlation}b). However the distance corresponding to the maximal value increases with increasing $\Gamma_6$. In the following we will show that this distance can be considered as an effective blockade radius in this system.

\subsection{Blockade radius in the presence of TBD}
Without TBD and for large single photon detuning, the blockade radius is $R_{0}\simeq \sqrt[6]{C_{6}\left\vert\gamma_e+i\Delta_{d}\right\vert /\Omega _{d}^{2}}$~\cite{Gorshkov2,Firstenberg2}, which is a result of the competition between the linewidth in the Rydberg state and the vdW interaction~\cite{Liu2,Petrosyan2,Pritchard1,Pritchard2,Yan1,Yan2,Liu1,Liu3}. The excitation is blocked within a volume determined by the blockade radius $R_0$, where only one Rydberg atoms can be excited. 

When the TBD is present, we note that the non-Hermitian Hamiltonian of the system is obtained,
\begin{equation}
\hat{H}_{eff}=\hat{H}_{0}+\hbar \sum_{j>k}\left( \frac{C_{6}}{R_{jk}^{6}}-i%
\frac{\Gamma _{6}}{2R_{jk}^{6}}\right) \hat{\sigma}_{dd}^{j}\hat{\sigma}%
_{dd}^{k},
\label{EffectiveEqu}
\end{equation}
where the vdW interaction and TBD are grouped together. By treating the two terms as a complex interaction, and using the same argument as we derive $R_0$, we can define a characteristic radius $R_b$,
\begin{equation}
R_{b}\simeq \sqrt[6]{\left\vert 1-i\frac{\Gamma _{6}}{2C_{6}}\right\vert }%
R_{0}, \label{Rbeff}
\end{equation}%
which depends on both the vdW interaction and TBD.

This radius increases with the TBD rate $\Gamma_6$. In the strong dephasing limit $\Gamma _{6}\gg C_{6}$, it is fully determined by the dephasing rate, $R_{b}\sim \sqrt[6]{\Gamma _{6}/2C_{6}}R_{0}$. Importantly the radius $R_b$ is identical to the distance corresponding to the maximal correlation, as shown in Fig.~\ref{fig:correlation}b and c. Such results are similar to the derivation of the blockade radius in conventional Rydberg-EIT~\cite{Firstenberg2}. Hence we will treat $R_b$ as an effective blockade radius for this dissipative optical medium.

\subsection{Enhancement of the blockade effect}
As the blockade radius is increased by the TBD, the blockade effect is enhanced in a high density atomic gas. In a blockade volume, the atoms are essentially two-level atoms (in states $|g\rangle$ and $|e\rangle$). They behave as a superatom, which has three collective states~\cite{Liu2,Petrosyan2,Yan1,Yan2,Liu1,Liu3}. In a homogeneous gas, we obtain the collective ground state $\left\vert
G\right\rangle =\left\vert g_{1},\cdots ,g_{N_{a}}\right\rangle $, singly
excited states $\left\vert E\right\rangle =\sum_{j}\left\vert g_{1},\cdots
,e_{j},\cdots ,g_{N_{a}}\right\rangle /\sqrt{N_{a}}$ and $\left\vert
D\right\rangle =\sum_{j}\left\vert g,\cdots ,d_{j},\cdots
,g_{N_{a}}\right\rangle /\sqrt{N_{a}}$ [see Fig.~\ref{fig1}(b)]. The
number $N_{a}$ of the blocked atoms in the volume $V=4\pi R_{b}^{3}/3$ of a superatom is given by $N_{a}=4\pi \rho R_{b}^{3}/3$, where $\rho $ is the density of the atomic gas. Hence the TBD increases the "mass" (i. e. the number of atoms) of a superatom (see Fig.~\ref{fig1}c and Fig.~\ref{fig:correlation}d). In the weak probe field limit, other states are prohibited from the dynamics due to the blockade.

In the one dimensional case, the number of
superatoms $N_{SA}=L/R_{b}$ reduces as the blockade radius increases. However the number of atoms that are blocked ${N}_{\text{tot}}=N_{SA}N_{a}=4\pi L\rho R_{b}^{2}/3$, which increases with increasing blockade radius. Therefore we obtain less superatoms, while the total number of blocked atoms (i.e. two-level atoms) is increased. These two-level atoms
breaks the EIT condition and causes light scattering.  As a result the transmission is reduced when the TBD rate is large. 

\section{Transmission and correlation of the probe light}
In this section, we will study stationary properties of the probe light in the presence of the vdW interaction and TBD. This will be done in the weak field limit through the Heisenberg-Langevin approach. We will work in the continuous limit,
which is valid when the atomic density is high. The one dimensional regime is realized when widths of light pulses are smaller
than the blockade radius.

\subsection{Heisenberg-Langevin equations}
Using the superatom model and the master equation (\ref%
{MasterEqu}) we obtain Heisenberg-Langevin equations of light and atomic operators in the
weak probe limit~\cite{Petrosyan2}
\begin{eqnarray}
\partial _{t}\hat{\mathcal{E}}_{p}\left( z\right) &=&-c\partial _{z}\hat{
\mathcal{E}}_{p}\left( z\right) +i\eta N\hat{\sigma}_{ge}\left( z\right) ,
\notag \\
\partial _{t}\hat{\sigma}_{ge}\left( z\right) &=&-\left( i\Delta _{p}+\gamma
_{e}\right) \hat{\sigma}_{ge}\left( z\right) -i\hat{\Omega _{p}^{\dag }}%
\left( z\right) -i\Omega _{d}\hat{\sigma}_{gd}\left( z\right) ,  \notag \\
\partial _{t}\hat{\sigma}_{gd}\left( z\right) &=&-i\left[ \Delta +\hat{S}%
_{V}\left( z\right) -i\hat{S}_{\Gamma }\left( z\right) \right] \hat{\sigma}%
_{gd}\left( z\right)  \notag \\
&&-\gamma _{d}\hat{\sigma}_{gd}\left( z\right) -i\Omega _{d}\hat{\sigma}%
_{ge}\left( z\right) ,  \label{HLEqu}
\end{eqnarray}%
where $\Delta =\Delta _{p}+\Delta _{d}$ is two-photon detuning. $\hat{S}%
_{V}\left( z\right) =\int d^{3}z^{\prime }\rho \left( z^{\prime }\right)
C_{6}/\left\vert z-z^{\prime }\right\vert ^{6}\hat{\sigma}_{dd}\left(
z^{\prime }\right) $ and $\hat{S}_{\Gamma }\left( z\right) =\int
d^{3}z^{\prime }\rho \left( z^{\prime }\right) C_{6}^{\prime }/2\left\vert
z-z^{\prime }\right\vert ^{6}\hat{\sigma}_{dd}\left( z^{\prime }\right) $
are spatially dependent interaction energy and TBD rate, respectively. Both
 $\hat{S}_V$ and $\hat{S}_{\Gamma}$ are nonlocal in the sense that these
quantities depend on the overall density $\rho(z)$ of the atomic gas and Rydberg probability operator $\hat{\sigma}_{dd}(z)$.
\begin{figure}[htbp]
	\centering
	\fbox{\includegraphics[width=0.95\linewidth]{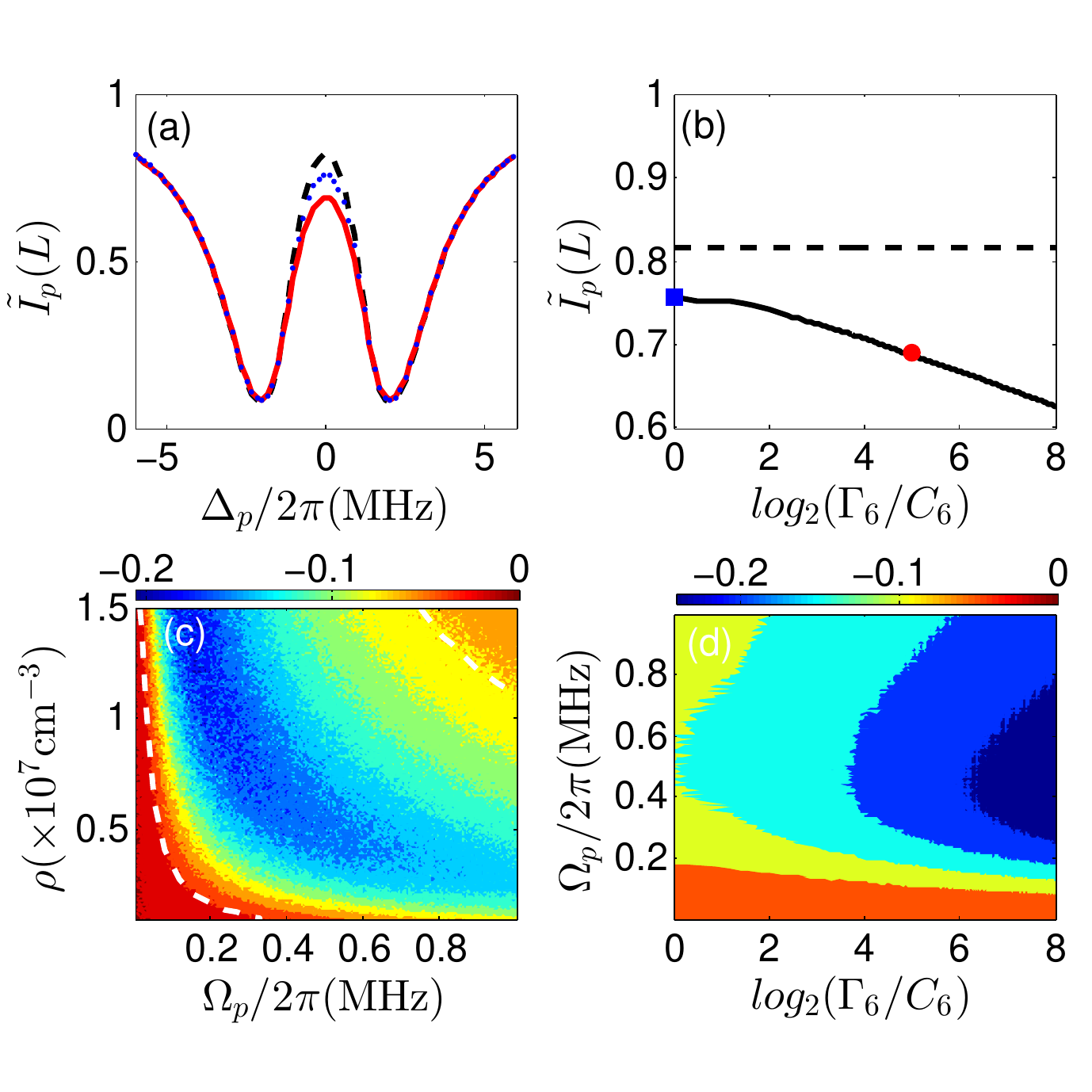}}
	\caption{(Color online) (\textbf{a}) Transmission v.s. the
		detuning $\Delta _{p}$ for TBD rate $\Gamma_6=0$ (dashed), $%
		\Gamma_6 = C_6$ (dotted) and $\Gamma_6=32C_6$ (solid). (\textbf{b})
		Dependence of the transmission on the TBD rate $\Gamma_6$ at the EIT
		resonance. The square and circle denote values of the transmission in (a)
		when $\Gamma_6=0$ and when $\Gamma_6=32C_6$. (\textbf{c}) Diagram of the transmission as a function of Rabi frequency
		$\Omega_p$ and atomic density $\protect\rho$. A TBD active region is found
		when $|\protect \delta \tilde{I}_{p}(L)|> 1\%$ (dashed line). The probe detuning $\Delta
		_{p}=0$. (\textbf{d}) Diagram of the transmission as a function of TBD rate $\Gamma_6$ and Rabi frequency $\Omega_p$. Increasing $\Gamma_6$ and $\Omega_p$ will reduce
		the transmission. The latter is caused by stronger blockade
		effect due to vdW density-density interactions. In panels (a), (b) and (d),
		the atomic density is $\protect\rho =0.5\times 10^{11}$ mm$^{-3}$. Rabi
		frequency $\Omega _{p}(0)/2\protect\pi =0.3$ MHz in (a) and (b). $%
		\Gamma_6=32C_6$ in panel (c). Other parameters are same with that of Fig.~\ref{fig:correlation}.}
	\label{fig:transmission}
\end{figure}

Knowing the blockade radius, we solve the
Heisenberg-Langevin equations of independent SAs in the steady state and
obtain the Rydberg excitation projection operator~\cite{Petrosyan2},
\begin{equation}
\hat{\Sigma}_{DD}\left( z\right) =\frac{N_{a}\eta ^{2}\hat{\mathcal{E}}%
_{p}^{\dag }\left( z\right) \hat{\mathcal{E}}_{p}\left( z\right) \Omega
_{d}^{2}}{N_{a}\eta ^{2}\hat{\mathcal{E}}_{p}^{\dag }\left( z\right) \hat{%
\mathcal{E}}_{p}\left( z\right) \Omega _{d}^{2}+\left( \Omega
_{d}^{2}-\Delta \Delta _{p}\right) ^{2}+\Delta ^{2}\gamma _{e}^{2}}.
\label{Sdd}
\end{equation}%
The polarizability of the probe field is conditioned on the projection,
\begin{equation}
\hat{P}\left( z\right) =\hat{\Sigma}_{DD}\left( z\right) P_{2}+\left[ 1-\hat{%
\Sigma}_{DD}\left( z\right) \right] P_{3}  \label{ConP}
\end{equation}%
where the polarizability becomes that of two-level atoms in a SA
\begin{equation}
P_{2}=\frac{i\gamma _{e}}{\gamma _{e}+i\Delta _{p}}  \label{P2}
\end{equation}%
and that of three-level atoms otherwise%
\begin{equation}
P_{3}=\frac{i\gamma _{e}}{\gamma _{e}+i\Delta _{p}+\frac{\Omega _{d}^{2}}{%
\gamma _{d}+i\Delta }}.  \label{P3}
\end{equation}%
It is clearly that optical response of a SA depends on the Rydberg
projection operator (\ref{Sdd}), i.e., SAs behave like a two-level, absorptive medium
due to $\hat{\Sigma}_{DD}\left( z\right) =1$.

The transmission of the probe light is captured by the probe light intensity
$I_{p}\left( z\right) =\langle \hat{\mathcal{E}}_{p}^{\dag }(z)\hat{\mathcal{%
E}}_{p}(z)\rangle $. In the steady state, the intensity $I_{p}\left(
z\right) $ satisfies a first order differential equation,
\begin{equation}
\partial _{z}\langle \hat{\mathcal{E}}_{p}^{\dag }(z)\hat{\mathcal{E}}%
_{p}(z)\rangle =-\kappa (z)\langle \mathrm{Im}[\hat{P}\left( z\right) ]\hat{%
\mathcal{E}}_{p}^{\dag }(z)\hat{\mathcal{E}}_{p}(z)\rangle ,  \label{PorEq1}
\end{equation}%
where $\kappa (z)=\rho \left( z\right) \omega _{p}/\left( \hbar \epsilon
_{0}c\gamma _{e}\right) $ denotes the resonant absorption coefficient.
Similarly we find the two-photon correlation function $g_{p}\left( z\right)
= $ $\langle \hat{\mathcal{E}}_{p}^{\dag 2}(z)\hat{\mathcal{E}}%
_{p}^{2}(z)\rangle /\langle \hat{\mathcal{E}}_{p}^{\dag }(z)\hat{\mathcal{E}}%
_{p}(z)\rangle ^{2}$ obeys~\cite{Petrosyan2}
\begin{equation}
\partial _{z}g_{p}(z)=-\kappa (z)\mathrm{Im}[P_{2}-P_{3}]\langle \hat{\Sigma}%
_{DD}(z)\rangle g_{p}(z).  \label{PorEq2}
\end{equation}%
The blockade radius is encoded in the correlation function of photon pairs,
which decays with the rate proportional to the excitation probability $%
\langle \hat{\Sigma}_{{DD}}\rangle$ and absorption rate of a
two-level atom when photon separation is smaller than the blockade radius.

To solve Eq.~(\ref{HLEqu})-(\ref{PorEq2}) the 1D atomic medium is divide
into $N_{SA}=L/\left( 2R_{b}\right) $ superatoms, and then we judge Rydberg
excitation whether $\left\langle \hat{\Sigma}_{DD}\left( z\right)
\right\rangle \rightarrow 1$ or $\left\langle \hat{\Sigma}_{DD}\left(
z\right) \right\rangle \rightarrow 0$ in each SA one by one via a Monte Carlo sampling. This procedure is repeated many times in order to evaluate mean values.

\subsection{Transmission of the probe field}
\label{sec:transmission}
The transmission of the probe field is characterized by the
ratio of light intensities at the output and input, i.e. $\tilde{I}%
_p(L)=I_{p}\left( L\right) /I_{p}\left( 0\right)$ with input values $%
I_{p}\left( 0\right) $. Without vdW interactions or TBD, high transmission is obtained in the EIT window $|\Delta_p|\le |\Omega_d|^2/\gamma_e$ due to the formation of dark state polaritons~\cite{Fleischhauer}. In the presence of the vdW interaction, the transmission is reduced due to the blockade effect. When turning on
the TBD, the transmission is further suppressed in
the EIT window, see Fig.~\ref{fig:transmission}a. Increasing the TBD strength $\Gamma_6$, the transmission $\tilde{I}_p(L)$ decreases gradually (Fig.~\ref%
{fig:transmission}b). A weaker transmission indicates that there are more atoms blocked
from forming dark state polaritons~\cite{Fleischhauer}. This is consistent with the analysis in Sec.~\ref{sec:blockade}B.

Outside the EIT window $|\Delta _{p}|>\Omega _{p}$, the transmission first
decreases with increasing detuning $\Delta _{p}$. It arrives at the minimal
transmission around the Autler-Townes splitting $\Delta _{p}=\pm \Omega _{d}$%
. In this region, the TBD is less important, and the transmission is almost
identical to cases when $\Gamma _{6}=0$ (Fig.~\ref{fig:transmission}a). Similar to the
transmission of EIT in a Rydberg medium~\cite{Petrosyan2}, the medium enters
a linear absorption regimes, where neither vdW interactions nor TBD affect
photon absorption dramatically.

In the following, we will focus on the transmission in the EIT window and
explore how the TBD interplays with other parameters. We first calculate the
transmission by varying atomic density and probe field Rabi frequency. To
highlight effects due to the TBD, we calculate differences of the
transmission with and without TBD, $\delta\tilde{I} = \tilde{I}_p(L)-\tilde{I%
}^{0}_p(L)$ where $\tilde{I}^{0}_p(L)$ denotes the light transmission when
the TBD is turned off. The result is shown in Fig.~\ref{fig:transmission}c. We find that
stronger probe field (larger $\Omega_p$) and higher atomic densities in
general lead to more pronounced TBD effect. The
``phase diagram'' shown in Fig.~\ref{fig:transmission}c allows us to distinguish TBD dominated regions. To do so, we plot a
phase boundary (dashed curve) when the difference $\delta\tilde{I}>1\%$.
Below this curve the transmission is largely affected by the vdW
interactions while above this curve, the atomic gas exhibits active TBD
phase. Namely, the transmission is reduced significantly due to the TBC.

In Fig.~\ref{fig:transmission}d, we show the
transmission by varying both the Rabi frequency $\Omega_p$ and TBD rate $%
\Gamma_6$. Fixing $\Gamma_6$, the transmission decreases with increasing $%
\Omega_p$. This results from the strong energy
shift caused by the vdW interaction~\cite{Petrosyan2,Gorshkov2}. On the other hand, the
transmission decreases with increasing $\Gamma_6$ if one fixes $\Omega_p$, i.e. the EIT is dominantly affected by the TBD. 

\begin{figure}[htbp]
	\centering
	\fbox{\includegraphics[width=0.95\linewidth]{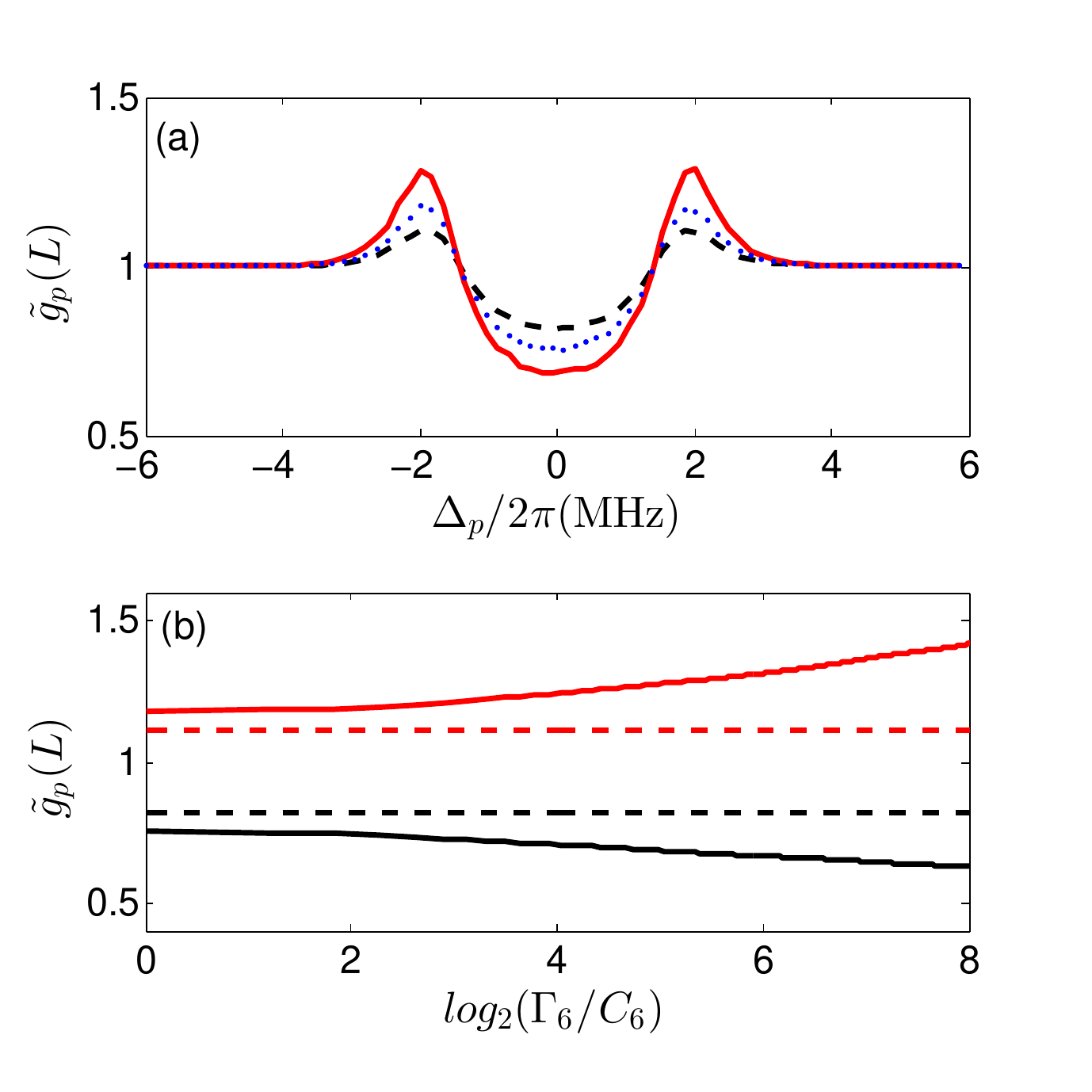}}
	\caption{(Color online) (\textbf{a}) Second-order correlation function $%
		\tilde{g}_{p}(L)$ versus the probe detuning $\Delta _{p}/2\protect\pi $ for
		TBD rate $\Gamma _{6}=0$ (dashed), $\Gamma _{6}=C_{6}$ (dotted) and $\Gamma
		_{6}=32C_{6}$ (solid). (\textbf{b}) Dependence of the second-order
		correlation function $\tilde{g}_{p}(L)$ on the TBD rate $\Gamma _{6}$ when $%
		\Delta _{p}/2\protect\pi =0.0$ MHz (black solid) and $\Delta _{p}/2\protect%
		\pi =2.0$ MHz (red solid). The dashed black curve ($\Delta _{p}/2\protect\pi %
		=0.0$ MHz) and dashed red curve ($\Delta _{p}/2\protect\pi =2.0$ MHz) denote
		the TBD rate $\Gamma _{6}=0$. Other parameters are the same as in Fig. 2.}
	\label{fig4}
\end{figure}
\subsection{Photon-photon correlation}
\label{sec:correlation}

The photon-photon correlation function exhibits nontrivial dependence on the TBD. The normalized correlation function $\tilde{g}%
_p(L)=g_{p}\left( L\right) /g_{p}\left( 0\right) $ at the exist of the
medium is shown in Fig.~\ref{fig4}a. In the EIT window, the correlation $\tilde{g}_p(L)$
becomes smaller when we turn on the TBD. Increasing the TBD strength $\Gamma_6$, the correlation decrease (see Fig.~\ref{fig4}a and Fig.~\ref{fig4}%
b). A smaller correlation indicates that anti-bunching becomes stronger. It is interesting to note that the transmission is large (Fig.~\ref{fig:transmission}a) in the EIT window. 

In contrast, the correlation $\tilde{g}_p(L)$ is enhanced
by the TBD outside the EIT window, $|\Delta_p|>\Omega_d^2/\gamma_e$. We
obtain maximal values of the correlation function around the Autler-Townes
doublet $\Delta_p\approx \pm\Omega_d$. Increasing $\Gamma_6$, the
maximal value (bunching) is also increased (see Fig.~\ref{fig4}c). We shall point out that
the transmission is smallest at the Autler-Townes doublet. It might become difficult to observe the TBD amplified bunching in this case, as the photon flux is low.

\section{Conclusions}
In summary, we have studied EIT in a one-dimensional gas of cold atoms involving highly excited Rydberg states. In this
model, each pair of atoms does not only experience the long-range vdW
interactions but also the nonlocal two-body dephasing. We show that the TBD
can enlarge the effective blockade radius. We demonstrate that in the EIT window, the TBD enhances the blockade effect, i.e. reducing the transmission and increasing photon-photon anti-bunching.
Away from the EIT window, the transmission is hardly affected by the TBD.
However, the photon bunching is amplified around the
Autler-Townes doublet.

Our work opens new questions in the study of Rydberg EIT. In the present
work, we focused on stationary states of extremely long light pulses at zero
temperature. It is worth studying how the combination of TBD and vdW
interactions will affect propagating dynamics of short light pulses. Moreover, it was found that the molecular coupling can cause non-stationary light transmission~\cite{Tresp}, while our model can not capture this feature. It is worth to developing new effective models to describe the transient dynamics in the future.

\begin{acknowledgements}
	D.Y acknowledges support from the National Natural Science Foundation of China (NSFC) Grants No. 11204019 and No. 11874004, the "Spring Sunshine" Plan Foundation of Ministry
	of Education of China Grant No. Z2017030, the Science Foundation of the Education Department of Jilin Province No.JJKH20200557KJ, and the China Scholarship Council (CSC) Grant No. 201707535012. Z.B acknowledges support from the NSFC Grant No. 11847221, the Shanghai Sailing Program Grant No. 18YF1407100, the China Postdoctoral Science Foundation Grant No. 2017M620140 and the International Postdoctoral Exchange Fellowship Program Grant No. 20180040. W.L. acknowledges support from the UKIERI-UGC Thematic Partnership No. IND/CONT/G/16-17/73, EPSRC Grant No. EP/M014266/1 and EP/R04340X/1, and support from the University of Nottingham.
\end{acknowledgements}

\appendix
\section*{Derivation of the two-body dephasing operator}
We consider a pair of atoms in Rydberg $|d\rangle$ state couple to a different Rydberg state $|r\rangle$ through a molecular process. This is described by the Hamiltonian $H_t=H + H_m$, where $H$ is the Hamiltonian given by Eq.~(\ref{Ham}), and the molecular Hamiltonian $H_m$ describes the dipolar interaction between the Rydberg states,
\begin{equation}
\hat{H}_m =  U(R_{12}) (\hat{\sigma}_{dr}^1\hat{\sigma}_{dr}^2 + \hat{\sigma}_{rd}^1\hat{\sigma}_{rd}^2),
\end{equation}
with the dipolar interaction $U(R_{12})=C_3/R_{12}^3$. Moreover the state $|r\rangle$ decays to the $|d\rangle$ through a single body spontaneous process. The dynamics is given by the master equation,
\begin{eqnarray}
\dot{\hat{\rho}}_m &=&-i[\hat{H}_m,\hat{\rho}_m] \\
&+&\gamma_r\sum_{j,k=1,2,j\neq k}\left(\hat{\sigma}_{dr}^j\hat{\rho}_m\hat{\sigma}_{rd}^k -\frac{1}{2}\{\hat{\rho}_m,\hat{\sigma}_{rd}^k\hat{\sigma}_{dr}^j\}\right). \nonumber
\end{eqnarray}
In the master equation, we assume that single body decay $\gamma_r$ is large and the molecular coupling is strong. The even weaker Hamiltonian $H$ will be taken into account adiabatically. 

We first focus on subspaces expanded by the two Rydberg states.  For strong single body decay, the system rapidly reaches to the equilibrium state. To consider different time scales, the master equation $\dot{\hat{\rho}} = (\mathcal{L}_0 +\mathcal{L}_1)\hat{\rho}$ is split into the fast (denoted by $\mathcal{L}_0\hat{\rho}$) and slow (denoted by $\mathcal{L}_1\hat{\rho}$) parts, where
\begin{eqnarray}
\frac{\mathcal{L}_0\hat{\rho}}{\gamma_r} &=& \sum_{j,k=1,2,j\neq k}\left(\hat{\sigma}_{dr}^j\hat{\rho}_m\hat{\sigma}_{rd}^k -\frac{1}{2}\{\hat{\rho}_m,\hat{\sigma}_{rd}^k\hat{\sigma}_{dr}^j\}\right), \nonumber\\
\mathcal{L}_1\hat{\rho} &=& -i[\hat{H}_m,\hat{\rho}_m].
\end{eqnarray}
We will trace the fast dynamics and derive an effective master equation for the slow dynamics via the second order perturbation calculation~\cite{gardiner04}.

Here we define a projection operator $\mathcal{P}_0=\lim_{t\to\infty}e^{t\mathcal{L}_0}$, which projects the density matrix to the subspace corresponding to the relatively slow dynamics, i.e. $\hat{\rho}=\mathcal{P}_0\hat{\rho}_m$. The first order correction vanishes $\mathcal{P}_0\mathcal{L}_1\mathcal{P}_0\hat{\rho}_m=0$. We then calculate the second order correction $-\mathcal{P}_0\mathcal{L}_1(\mathcal{I}-\mathcal{P}_0)\mathcal{L}_1\mathcal{P}_0\hat{\rho}_m$. A tedious but straightforward calculation yields an effective master equation depending on the two-atom dephasing,
\begin{equation}
\hat{\rho}_e \approx \frac{2U^2(R_{12})}{\gamma_r}\left(\hat{\sigma}_{dd}^1\hat{\sigma}_{dd}^2\hat{\rho}_e\hat{\sigma}_{dd}^2\hat{\sigma}_{dd}^1 -\frac{1}{2}\{\hat{\sigma}_{dd}^2\hat{\sigma}_{dd}^1,\hat{\rho}_e\}\right).
\end{equation}
Defining $\Gamma_{12} = 2U^2(R_{12})/\gamma_r$ and taking Hamiltonian $H$ and other process into account adiabatically, we obtain the master equation given in the main text (by further extending the approximate result to the many-atom setting).


\end{document}